\documentclass[twocolumn,trackchanges]{aastex62}
\usepackage{booktabs}
\usepackage{color}
\usepackage{soul}
\usepackage{ulem}
\usepackage{bm}
\usepackage{amsmath}

\hypersetup{linkcolor=red,citecolor=blue,filecolor=cyan,urlcolor=magenta}

\submitjournal{ApJS}

\shortauthors{Wang et al.}

\begin{document}

\title{Large-scale multiconfiguration Dirac-Hartree-Fock calculations for astrophysics: Cl-like ions from  Cr~VIII to Zn~XIV}

\author{K. Wang}
\affil{Hebei Key Lab of Optic-electronic Information and Materials, The College of Physics Science and Technology, Hebei University, Baoding 071002, China;}
\affil{Chimie Quantique et Photophysique, CP160/09, Universit\'{e} libre de Bruxelles, Av. F.D. Roosevelt 50, 1050 Brussels, Belgium;}
\affil{Shanghai EBIT Lab, Key Laboratory of Nuclear Physics and Ion-beam Application, Institute of Modern Physics, Department of Nuclear Science and Technology, Fudan University, Shanghai 200433, China;}

\author{P. J\"onsson}
\affil{Group for Materials Science and Applied Mathematics, Malm\"o University, SE-20506, Malm\"o, Sweden; \\}

\author{G. Del Zanna}
\affil{DAMTP, Centre for Mathematical Sciences, University of Cambridge, Wilberforce Road, Cambridge CB3 0WA, UK}

\author{M. Godefroid}
\affil{Chimie Quantique et Photophysique, CP160/09, Universit\'{e} libre de Bruxelles, Av. F.D. Roosevelt 50, 1050 Brussels, Belgium;}

\author{Z. B. Chen}
\affil{Department of Applied Physics, School of Science, Hunan University of Technology, Zhuzhou 412007, China;  chenzb008@qq.com}

\author{C. Y. Chen}
\affil{Shanghai EBIT Lab, Key Laboratory of Nuclear Physics and Ion-beam Application, Institute of Modern Physics, Department of Nuclear Science and Technology, Fudan University, Shanghai 200433, China;}
\author{J. Yan}
\affil{Institute of Applied Physics and Computational Mathematics, Beijing 100088, China;}
\nocollaboration



\begin{abstract}
We use the multiconfiguration Dirac-Hartree-Fock (MCDHF) method combined with the relativistic configuration interaction (RCI) approach
(GRASP2K) to  provide  a consistent  set of transition energies and radiative transition data for the lower  $n =3$ states
 in all Cl-like ions of astrophysical importance, from \ion{Cr}{8}  to \ion{Zn}{14}.
 We also provide excitation energies calculated for \mbox{Fe X} using the many-body perturbation theory (MBPT, implemented within FAC). 
The comparison of the present MCDHF results with MBPT and with the available experimental energies indicates that 
the theoretical excitation energies are highly accurate, with uncertainties of only a few hundred cm$^{-1}$.  
Detailed comparisons for  \ion{Fe}{10} and \ion{Ni}{12} highlight 
discrepancies in the experimental energies found in the literature. Several 
new identifications are proposed. 
 \end{abstract}

\keywords{atomic data - atomic processes}



\section{Introduction} \label{sec:intro}

Cl-like ions produce several strong transitions that have been widely used in 
astrophysics for a range of diagnostic applications, to e.g. 
measure electron temperatures, densities, chemical composition and even strong magnetic fields~\citep{Zanna.2012.V541.p90}. 
For a review of some of the applications, see \cite{delzanna_mason:2018}. 
Providing accurate atomic data for these ions is of paramount importance.

Using the multiconfiguration Dirac-Hartree-Fock (MCDHF) and the relativistic configuration interaction (RCI) methods \citep{FroeseFischer.2016.V49.p182004,Grant.2007.V.p},
we  provide here  a consistent  set of transition energies and radiative transition data with high accuracy in all Cl-like ions of astrophysical importance, from \ion{Cr}{8} ($Z=24$)  to \ion{Zn}{14} ($Z=30$). 
Energy levels, wavelengths, oscillator strengths, line strengths, transition rates, and lifetimes for the main $n=3$ levels of the $3s^2 3p^5$, $3s 3p^6$, $3s^2 3p^4 3d$,  $3s 3p^5 3d$, $3s^2 3p^3 3d^2$, $3s 3p^4 3d^2$, and $3s 3p^3 3d^3$ configurations are provided.  To assess the accuracy of the MCDHF transition energies, we have also performed  calculations and provided excitation energies for \mbox{Fe X} using the many-body perturbation theory (MBPT)~\citep{Lindgren.1974.V7.p2441}. This work extends and complements our long-term theoretical efforts~\citep{Wang.2014.V215.p26,Wang.2015.V218.p16,Wang.2016.V223.p3,Wang.2016.V226.p14,Wang.2017.V119.p189301,Wang.2017.V194.p108,Wang.2017.V187.p375,Wang.2017.V229.p37,Wang.2018.V235.p27,Wang.2018.V239.p30,Wang.2018.V234.p40,Wang.2018.V208.p134,Wang.2019.V236.p106586,Chen.2017.V113.p258,Chen.2018.V206.p213,Guo.2015.V48.p144020,Guo.2016.V93.p12513,Si.2016.V227.p16,Si.2017.V189.p249,Si.2018.V239.p3,Zhao.2018.V119.p314} to provide atomic data for L- and M-shells systems with high accuracy.
For a review, see \cite{Jonsson.2017.V5.p16}.

In Section~2 we briefly describe the calculations, while in Section~3 we present comparisons 
between theoretical and experimental energies for the low-lying levels of the main ions. We discuss in 
some detail the energies of the two main ions, \ion{Fe}{10} and \ion{Ni}{12}, 
revising some previous identifications. We then present our transition rates.

\section{Theory and Calculations}
The MCDHF method in the GRASP2K code~\citep{Jonsson.2013.V184.p2197,Jonsson.2007.V177.p597} and the MBPT method in the FAC code~\citep{Gu.2008.V86.p675,Gu.2007.V169.p154}
 are described by~\citet{FroeseFischer.2016.V49.p182004} and by ~\citet{Lindgren.1974.V7.p2441}, respectively. These two methods are also introduced in our recent papers~\citep{Wang.2018.V235.p27,Wang.2018.V239.p30}. For this reason, in the sections below, only the computational procedures are described.

\subsection{MCDHF}\label{Sec:MCDHF}
In our MCDHF calculations, the multireference (MR) sets for even and odd parities include
\begin{itemize}
\item [even:]
$3s 3p^6$, $3s^2 3p^4 3d$, $3s 3p^4 3d^2$,  $3p^6 3d$, $3s^2 3p^2 3d^3$, $3p^4 3d^3$, $3s 3p^2 3d^4$, $3s 3p^5 4p$, $3s^2 3p^4 4s$;
\item [odd:]
$3s^2 3p^5$, $3s 3p^5 3d$, $3s^2 3p^3 3d^2$,  $3p^5 3d^2$, $3s 3p^3 3d^3$, $3s^2 3p 3d^4$, $3s^2 3p^4 4p$, $3s 3p^5 4s$.
\end{itemize}

Initial MCDHF calculations for the MR sets for even and odd parities are performed to determine simultaneously all the orbitals $1s$, $2s$, $2p$, $3s$, $3p$, $3d$, $4s$, and $4p$ of the MR sets. 
Then, by allowing single  and double  substitutions from the $3s$, $3p$, $3d$, $4s$, $4p$ electrons of the MR sets to  orbitals with $n\leq7, l\leq5$, and single excitations of the $2s$ and $2p$ electrons to orbitals with $n\leq6, l\leq5$,  configuration state function (CSF) expansions are obtained. We consider the $1s$ shell as inactive, keeping its two electrons in all CSFs of the expansions. In order to monitor and obtain the convergence of computed properties, the orbital set is  increased systematically layer by layer. At each stage only the outer orbitals are optimized, while the inner ones are fixed.
Both valence-valence (VV) electron correlation effects and core-valence (CV) electron correlation effects are required for obtaining accurate results, though VV electron correlation effects are more important. Among CV electron correlation effects arising from subshells $2s$ and $2p$, those from  $2p$ are the most important.  We also checked that the contributions  involving orbitals with principal quantum number $n=8$ for VV electron correlation are negligible and that opening the $1s$ shell for CV electron correlation can be omitted.

Once the orbitals optimized, the  QED corrections and Breit interaction are included in the RCI calculations. 
The numbers of CSFs in the final  even and odd expansions, are, respectively, around  26 and 23 millions, for the two parities.
We use the $jj$-$LSJ$ transformation approach~\citep{Gaigalas.2017.V5.p6,Gaigalas.2004.V157.p239} to transform the  $jj$-coupled CSFs into $LSJ$-coupled CSFs, and obtain the $LSJ$ labels used by experimentalists.

\subsection{MBPT}
The MBPT method~\citep{Lindgren.1974.V7.p2441} is implemented in the FAC code by~\citet{Gu.2008.V86.p675,Gu.2007.V169.p154}. This method was used in our recent papers
~\citep{Wang.2014.V215.p26,Wang.2015.V218.p16,Wang.2016.V223.p3,Wang.2016.V226.p14,Wang.2017.V229.p37,Wang.2018.V239.p30}
to provide high accuracy atomic data for L- and M- shell ions.
In the  MBPT method, the Hilbert space of the full Hamiltonian is divided into two parts, i.e., a model space $M$ and a orthogonal space $N$. 
We included in $M$ all the CSFs of the MR set as defined above for the MCDHF calculations. All the possible CSFs generated by allowing single  and double  substitutions from the electrons of the MR sets are included in the $N$ space. 
With the maximum $l$-value of 20, the maximum $n$-values considered for the single and double  excitations are, respectively, 125 and 65. 
The configuration interaction effects in the model space $M$ are considered non-perturbatively, i.e. are included in the self-consistent field calculations. 
The interaction effects between the two spaces $M$ and $N$ are considered through the second-order perturbation theory.



\section{EVALUATION OF DATA}

\subsection{Energy Levels and Lifetimes}~\label{Sec:en}

Given its importance, we focus first on \ion{Fe}{10}. The energies of the $3s^2 3p^5$, $3s 3p^6$
levels were well established, whilst not all of the $3s^2 3p^4 3d$ levels (producing the strongest lines)
were identified in the earlier studies in the 1960's and 1970's. For a review on the accuracy of the 
experimental energies see \cite{DelZanna.2004.V422.p731},
 where several values were improved, and alternative/new
identifications were proposed. \cite{DelZanna.2004.V422.p731} used the {\sc superstructure} code and applied,
whenever possible, Term Energy Corrections (TEC) to improve the theoretical energies. 
In a later scattering calculation for the $n=4$ levels, \citet{Zanna.2012.V541.p90} revised a 
few of the identifications using {\sc autostructure} \citep{badnell:2011}, which were included in the 
CHIANTI database~\citep{Dere.1997.V125.p149,Dere.2019.V241.p22}. 

Several structure calculations have been published over the years, and some comparisons are shown in 
Table~\ref{tab.lev.fe}.
In this Table, excitation energies for the lowest 197 states of the $3s^2 3p^5$, $3s 3p^6$, $3s^2 3p^4 3d$,  $3s 3p^5 3d$, $3s^2 3p^3 3d^2$, and $3s 3p^4 3d^2$ configurations in  \ion{Fe}{10} from the present MCDHF and MBPT calculations are given. Observed energies from the NIST database 
 and computed values from different sources are also included: CHIANTI, the MR-MP calculations by \citet{Ishikawa.2010.V43.p74022}, and  the {\sc grasp} calculations by~\citet{Aggarwal.2005.V439.p1215}. 
We note that   \citet{Ishikawa.2010.V43.p74022} and \citet{Aggarwal.2005.V439.p1215} 
provide results only for the lowest 90 states. 
The {\sc grasp} calculations by~\citet{Aggarwal.2005.V439.p1215} focus on the estimation of cross-sections for electron impact excitation for the lowest 90 levels of the  $3s^2 3p^5$, $3s3p ^6$, $3s^2 3p^4 3d$, $3s3p^5 3d$, and $3s^2 3p^3 3d^2$ configurations, and restricted electron correlation effects to the $n=3$ valence shells. In their calculations, valence-valence electron correlation effects among the levels of the $3s^2 3p^5$, $3s 3p^6$, $3s^2 3p^4 3d$, $3s 3p^5 3d$, $3p^6 3d$, $3s 3p^4 3d^2$, $3s^2 3p^3 3d^2$, and $3s^2 3p^2 3d^3$ configurations are included. Because of this limitation,  results from \citet{Aggarwal.2005.V439.p1215} are generally higher than our MCDHF values 
by one to several tens of thousand cm$^{-1}$.   
As can be seen in Table~\ref{tab.lev.fe}, good agreement is obtained between the MCDHF, MBPT and MR-MP calculations for the lowest 90 states.

For higher-lying levels (above $\#90$, mostly for the  $3s^2 3p^3 3d^2$ levels) the only theoretical results available for comparison are the {\sc autostructure}  calculations by~\citet{Zanna.2012.V541.p90}.  The latter focused on the estimation of cross-sections for electron impact excitation for the higher levels and were limited to the $n=4$ configurations, restricting electron correlation to the valence shells.
 Because of this limitation, the {\sc autostructure} excitation energies differ from our MCDHF results by $-$1~500 cm$^{-1}$ -- 22~000 cm$^{-1}$ for the higher-lying levels.  
On the other hand,  good agreement is obtained among the present accurate MCDHF and MBPT  calculations.

\begin{figure*}[ht!]
	\plotone{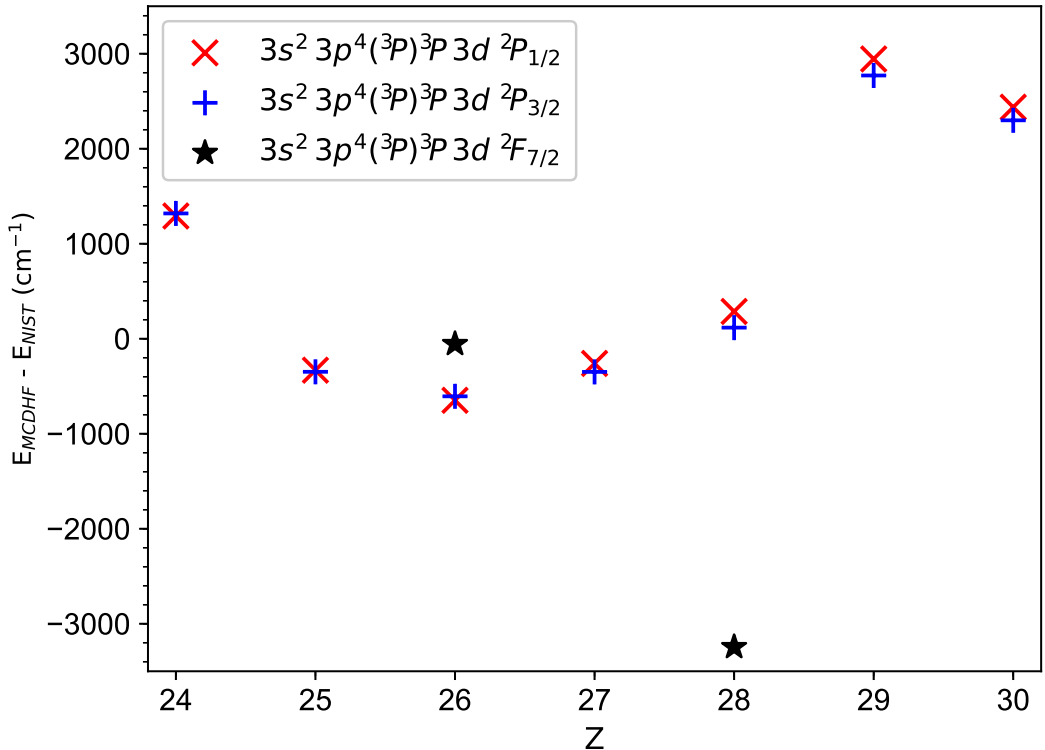}
	\caption{The differences (in cm$^{-1}$) of the MCDHF and NIST excitation energies for the $3s^{2}\,3p^{4}(^{3}P)~^{3}P\,3d~^{2}P_{1/2}$,  $3s^{2}\,3p^{4}(^{3}P)~^{3}P\,3d~^{2}P_{3/2}$ and $3s^{2}\,3p^{4}(^{3}P)~^{3}P\,3d~^{2}F_{7/2}$  levels as a function of $Z$. The data come from Table~\ref{tab.lev.all}.\label{figure.lev.nistlargedifferences}}
\end{figure*}

Based on our MCDHF and MBPT excitation energies, we confirm most of experimental values 
from the NIST ASD and the CHIANTI database, except for a few cases, which we now discuss. 
For the level $\#11/3s^{2}\,3p^{4}(^{3}P)~^{3}P\,3d~^{4}F_{5/2}$, our energy (426602 cm$^{-1}$)
is very close to the value suggested by \cite{jupen_etal:1993}, 426707  cm$^{-1}$, allowing us to confirm their identification.

For the state $\#12/3s^{2}\,3p^{4}(^{1}D)~^{1}D\,3d~^{2}P_{3/2}$, the NIST energy is clearly in error.
Our energy, 427951 cm$^{-1}$, is very close to the experimental value suggested by 
\cite{jupen_etal:1993}, 428002 cm$^{-1}$. We therefore confirm their identification.

For the level $\#16/3s^{2}\,3p^{4}(^{3}P)~^{3}P\,3d~^{4}P_{3/2}$, our MCDHF value is 439619 cm$^{-1}$, 
relatively close to the value suggested by \cite{DelZanna.2004.V422.p731} 
(438168 cm$^{-1}$), but not to the value later proposed by \citet{Zanna.2012.V541.p90}, and included in the  CHIANTI database, version 9. Our value is not close to any experimental energies previously suggested. 
\cite{jupen_etal:1993} suggested a value of 442439 cm$^{-1}$, definitely too far from our value. 
Considering the good agreement between our excitation energies and observed energies for the other $^{4}P$ levels, i.e., $3s^{2}\,3p^{4}(^{3}P)~^{3}P\,3d~^{4}P_{1/2,5/2}$, we believe that  the observed energy should be close (within 200 cm$^{-1}$) to our predicted one for the level $\#16/3s^{2}\,3p^{4}(^{3}P)~^{3}P\,3d~^{4}P_{3/2}$, and that the latest tentative identification in the  CHIANTI database should be discarded.

Our energy for the level $\#19/3s^{2}\,3p^{4}(^{1}D)~^{1}D\,3d~^{2}D_{5/2}$ is in good agreement with the 
experimental value (444127 cm$^{-1}$) suggested by \cite{jupen_etal:1993}.

For the state $\#22/3s^{2}\,3p^{4}(^{3}P)~^{3}P\,3d~^{2}F_{5/2}$ the NIST energy is clearly in error, while our energy 
supports the identification in  CHIANTI, due to \cite{DelZanna.2004.V422.p731}. 
For the state $\#23/3s^{2}\,3p^{4}(^{1}D)~^{1}D\,3d~^{2}F_{5/2}$, the experimental value 
in the CHIANTI database, suggested by \cite{DelZanna.2004.V422.p731} (and also \cite{jupen_etal:1993}),
 shows good agreement ($\Delta E_{\rm{CHIANTI}}$ = -41 cm$^{-1}$) with our MCDHF value, 
while the NIST values differs from our value by 5~388 cm$^{-1}$.
Our energy for level $\#26/3s^{2}\,3p^{4}(^{1}S)~^{1}S\,3d~^{2}D_{5/2}$ (516334 cm$^{-1}$) is 
very close to the value suggested by  \cite{jupen_etal:1993}, 516222~cm$^{-1}$.
 For the state $\#35/3s~^{2}S\,3p^{5}~^{3}P\,3d~^{4}F_{9/2}^{\circ}$, 
the experimental value from the NIST database 
(based on a suggestion rather than a firm observation) shows good agreement 
($\Delta E_{\rm{NIST}}$ = $-$180 cm$^{-1}$) with our MCDHF value, 
while the CHIANTI value, based on a tentative alternative suggestion, 
differs from our value by $-$5~052 cm$^{-1}$.

The above comparisons clearly demonstrate the importance of the present 
calculations to assess the correctness of level identifications.

With regard to the lifetimes in \ion{Fe}{10}, our MCDHF values, $\tau_{\rm MCDHF}^l$ in the length form and  $\tau_{\rm MCDHF}^v$ in the velocity form, 
show good agreement with an average deviation of 1.5~\%. 
Large deviations with the values for the lower levels calculated by \citet{Zanna.2012.V541.p90} are found.
This is not surprising as those calculations were aimed at the $n = 4$ configurations.

After \ion{Fe}{10}, the most important ion (considering its  abundance in astrophysical plasma) is 
 \ion{Ni}{12}. Compared to \ion{Fe}{10}, the status of the experimental energies for this ion is very poor, as described in 
detail by a recent  review of laboratory and astrophysical 
observations in \cite{delzanna_badnell:2016_ni_12}. In that paper, new scattering calculations were 
used to provide several  tentative identifications for the unknown energy levels. 
Their  {\sc autostructure} excitation energies are generally relatively close (less than 1~000 cm$^{-1}$) to our MCDHF
values, but we revise here many of the suggested experimental energies, see Table~\ref{tab.ni_12}. As described in  \cite{delzanna_badnell:2016_ni_12},
the energies of many of the $3d$ levels are known only relative to the $\#5/3s^{2}\,3p^{4}(^{3}P)~^{3}P\,3d~^{4}D_{7/2}$ metastable state. 
The identification provided by \cite{delzanna_badnell:2016_ni_12} (a decay to the ground state at 220.247~\AA)
 is close to the approximate value given by NIST, but is over  3~000~cm$^{-1}$ higher than our MCDHF energy. 
Considering the overall  accuracy of our calculations, we suggest here an alternative identification. The decay 
of this level could indeed coincide with an \ion{Fe}{13} transition at 221.822~\AA, providing an energy of  450812 cm$^{-1}$, only 129 cm$^{-1}$
lower than our MCDHF value. This identification was not considered in \cite{delzanna_badnell:2016_ni_12}
because the high-resolution solar spectrum of \cite{Behring.1976.V203.p521} did not suggest any line blending.
However, we have estimated the intensity of the \ion{Fe}{13} transition relative to those of other nearby 
lines using the atomic data from \cite{delzanna_storey:12_fe_13}, and found that only about half the observed intensity 
is due to \ion{Fe}{13}. The other half would be due to \ion{Ni}{12}.
We still regard this as a tentative new identification, as the decay from the $^{4}D_{5/2}$ 
level should be also observed. If the decay of the $^{4}D_{7/2}$ is at  221.822~\AA, the decay from the $^{4}D_{5/2}$  level 
should be at 222.35~\AA, but no line has been reported around this wavelength. 
Assuming that all these blends occur, we can  infer the energies of several levels 
($\#8$, $\#10$, $\#16$, $\#20$, $\#21$, $\#24$), shown in  Table~\ref{tab.ni_12}. We can see that all the inferred energies are very close to our 
ab initio  MCDHF values, thus giving us confidence in the new identifications. 
 
We also tentatively assign the energy of level $\#11$ ($^{4}F_{5/2}$) from a laboratory wavelength of 201.47~\AA, 
and that of level $\#17$ ($^{4}P_{3/2}$) from a laboratory wavelength of 195.51~\AA.
For the levels $\#19$, $\#22$, and $\#30$ we confirm the identifications proposed by \cite{delzanna_badnell:2016_ni_12}.
The identification of level $\#31$, the $^{2}D_{3/2}$, has been troublesome, as described in a section 
of the \cite{delzanna_badnell:2016_ni_12} paper. On the basis of the high-resolution laboratory spectrum 
of \cite{ryabtsev:1979} and the calculated intensities, \cite{delzanna_badnell:2016_ni_12} suggested that 
one of the two lines observed at 152.697 and  152.929~\AA\ should be the decay from the 
 $^{2}D_{3/2}$ level. The preference was given to the first one, as there is a line in the 
\cite{Behring.1976.V203.p521} solar spectrum at  152.703~\AA, although too strong. Here we prefer the 
second option, as it gives an energy closer to our MCDHF value.

Finally, a summary of excitation energies (in cm$^{-1}$) and radiative lifetimes (in s) for the lowest 187 (196, 197, 193, 192, 194, 194) states of the $3s^2 3p^5$, $3s 3p^6$, $3s^2 3p^4 3d$,  $3s 3p^5 3d$, $3s^2 3p^3 3d^2$, $3s 3p^4 3d^2$, and $3s 3p^3 3d^3$ configurations in  Cr~VIII (Mn~IX, Fe~X, Co~XI, Ni~XII, Cu~XIII, Zn~XIV) from the present MCDHF calculations are provided in  Table~\ref{tab.lev.all}. 
All the states  lie under the first $3p^6 3d$ state. 
Observed values compiled in the NIST ASD are also included in Table \ref{tab.lev.all}. The differences for observed values from the present MCDHF results are about 2 500 cm$^{-1}$ -- 3 200 cm$^{-1}$ in Ni~XII and Cu~XIII, and about 2 000 cm$^{-1}$ -- 2 500 cm$^{-1}$ in Zn~XIV.  Whereas good agreement is obtained for the same levels in lower-$Z$ ions (Mn~IX, Fe~X, and Co~XI). 
The differences (in cm$^{-1}$) between the MCDHF and NIST excitation energies for the $3s^{2}\,3p^{4}(^{3}P)~^{3}P\,3d~^{2}P_{1/2}$,  $3s^{2}\,3p^{4}(^{3}P)~^{3}P\,3d~^{2}P_{3/2}$ and $3s^{2}\,3p^{4}(^{3}P)~^{3}P\,3d~^{2}F_{7/2}$  levels, as a function of the  nuclear charge $Z$ are shown in Figure~\ref{figure.lev.nistlargedifferences} as an example. 
For $3s^{2}\,3p^{4}(^{3}P)~^{3}P\,3d~^{2}P_{1/2}$ and $3s^{2}\,3p^{4}(^{3}P)~^{3}P\,3d~^{2}P_{3/2}$ four anomalies appear in Cu~XIII and Zn~XIV. 
 Since in the present MCDHF calculations the same computational strategies are used for each ion, the accuracy of our calculated excitation energies  should be consistent and systematic for the same level along the sequence. 
Therefore, the large differences indicate that several other identifications need to be revised.

\subsection{Transition rates}
Wavelengths $\lambda_{ij}$, transition
rates $A_{ji}$, and branching fractions 
 (${\rm BF}_{ji} = A_{ji}/ \sum \limits_{k=1}^{j-1} A_{jk}$)
involving all levels considered in the present MCDHF calculations, as reported in Table~\ref{tab.lev.all}, along with line strength $S_{ji}$ and weighted oscillator strengths $gf_{ji}$, are provided in Table \ref{tab.trans.all}.  
E1 and E2 transition data in both length ($l$) and velocity ($v$) forms are given.  
For E1 and E2 transitions, we provide (last column) the uncertainty estimations of line strengths $S$ adopting the NIST ASD~\citep{Kramida.2018.V.p} terminology (A$^{+}$ $\leq$ 2~\%, A $\leq$ 3~\%, B$^{+}$ $\leq$ 7~\%, B $\leq$ 10~\%, C$^{+}$  $\leq$ 18~\%,  C $\leq$ 25~\%,  D$^{+}$ $\leq$ 40~\%, D $\leq$ 50~\%, and E $>$ 50~\% )   
and using the method proposed by~\cite{Kramida.2014.V212.p11}. 
For each E1 transition, the deviation $\delta S$  of line strengths $S_l$ in the length form  and $S_v$  in the velocity form  is defined as 
$\delta S$ = $|S_{v} - S_{l}|$/max($S_{v}$,~$S_{l}$). In various ranges of $S$, the averaged uncertainties $\delta S_{av}$ of $\delta S$ for E1 transitions in Fe~X  are assessed to 1.1~\% for $S \geq 10^{0}$; 1.3~\% for $10^{0} > S \geq 10^{-1}$; 2.2~\% for $10^{-1} > S \geq 10^{-2}$; 3.3~\% for $10^{-2} > S \geq 10^{-3}$; 7.3~\% for $10^{-3} > S \geq 10^{-4}$, 14~\% for $10^{-4} > S \geq 10^{-5}$, and 29~\% for $10^{-5} > S \geq 10^{-6}$. Then,  the largest of $\delta S_{av}$  and $\delta S_{ij}$ is considered to be the uncertainty of each particular transition. 

In Table~\ref{tab.trans.all}, about 8.4~\% E1 transitions in Fe~X have  the uncertainty of A+, 18.2~\% of A, 30.8~\%  of  B+, 18.1~\% of  B, 15.9~\%  of  C+, 2.4~\% of C, 4.3~\% have the uncertainty of  D+, and 1.2~\% of  D, while only 0.7~\% have the uncertainty of E. 

Using the same classification method, the uncertainties of the line strength $S$ for E2 transitions in Fe X, as well as those for E1 and E2 transitions in Cr~VIII, Mn~IX, Co~XI, Ni~XII, Cu~XIII, and Zn~XIV, are also listed in Table~\ref{tab.trans.all}.

\subsection{Summary}
The calculations of excitation energies, lifetimes, and radiative transition data for the  $n=3$ states of  Cl-like ions from  Cr~VIII to Cu~XIV were  performed using the MCDHF/RCI methods. 
Our detailed discussion of the energies of  \ion{Fe}{10} and \ion{Ni}{12} have highlighted several 
discrepancies in the experimental energies in NIST and in the literature. 
The above comparisons clearly show the importance of the present ab initio
calculations to assess the correctness of level and line identifications. Further studies are required to assess the identifications of the other ions, and further experimental work is encouraged 
to confirm our suggestions, especially for \ion{Ni}{12}.

\section*{Scientific software packages}
Scientific software packages including~\software{GRASP2K \citep{Jonsson.2007.V177.p597,Jonsson.2013.V184.p2197}
and	FAC \citep{Gu.2008.V86.p675}} are used in the present work. 	

\acknowledgments
We acknowledge the support from the National Key Research and Development Program of China under Grant No.~2017YFA0402300, the Science Challenge Project of China Academy of Engineering Physics (CAEP) under Grant No. TZ2016005, the National Natural Science Foundation of China (Grant No.~11703004, No.~11674066, No.~11504421, and No.~11474034), the Natural Science Foundation of Hebei Province, China (A2019201300 and A2017201165), and the Natural Science Foundation of Educational Department of Hebei Province, China (BJ2018058). This work is also supported by the Fonds de la Recherche Scientifique - (FNRS) and the Fonds Wetenschappelijk Onderzoek - Vlaanderen (FWO) under EOS Project n$^{\rm o}$~O022818F, and by the Swedish research council under contracts 2015-04842 and 2016-04185.  
GDZ acknowledges support form STFC (UK) via the consolidated grant to the solar/atomic astrophysics group, DAMTP, University of Cambridge.
KW expresses his gratitude to the support from the visiting researcher program at the Fudan University.

\clearpage
\bibliographystyle{aasjournal}
\bibliography{ref}

\clearpage
\startlongtable

\begin{deluxetable*}{clrrrrrrrrcc}
	\centering
	\tabletypesize{\tiny}
	\setlength{\tabcolsep}{3pt}
	\tablecaption{Excitation energies  $E$ in cm$^{-1}$ and radiative lifetimes $\tau$ in s for 197 states of \ion{Fe}{10}.  In addition to \st{Except for} the present MCDHF excitation energies $E_{\rm MCDHF}$, differences ($\Delta E_x = E_x - E_{\rm MCDHF}$) in cm$^{-1}$ from different sources (see below) are listed. $\Delta E_{\rm MBPT}$ - the present  MBPT calculations; $\Delta E_{\rm AS}$ -  the {\sc autostructure}  calculations by~\citet{Zanna.2012.V541.p90}; $\Delta E_{\rm NIST}$ - observed energies listed in the NIST ASD \citep{Kramida.2018.V.p}; $\Delta E_{\rm{CHIANTI}}$ - observed energies 
evaluated by  \cite{DelZanna.2012.V546.p97,DelZanna.2004.V422.p731} and listed in  the  CHIANTI database \citep{Dere.1997.V125.p149,Dere.2019.V241.p22}; $\Delta E_{\rm{MRMP}}$ - the MR-MP calculations by \citet{Ishikawa.2010.V43.p74022};  $\Delta E_{\rm{GRASP}}$ - the {\sc grasp} calculations by~\citet{Aggarwal.2005.V439.p1215};  $\tau_{\rm{MCDHF}}^l$ - the present MCDHF calculations in the length form;  $\tau_{\rm{MCDHF}}^v$ -  the present MCDHF calculations in the velocity form; $\tau_{\rm{AS}}^l$  - the {\sc autostructure}  calculations in the length form by~\citet{Zanna.2012.V541.p90}. 
		\label{tab.lev.fe}
	}
	\tablehead{\colhead{Key} & \colhead{Level} & \colhead{$E_{\rm MCDHF}$} & \colhead{$\Delta E_{\rm MBPT}$}& \colhead{$\Delta E_{\rm AS}$}  & \colhead{$\Delta E_{\rm NIST}$} & \colhead{$\Delta E_{\rm CHIANTI}$} & \colhead{$\Delta E_{\rm{MRMP}}$} &   \colhead{$\Delta E_{\rm{GRASP}}$} &  \colhead{$\tau_{\rm{MCDHF}}^l$} & \colhead{$\tau_{\rm{MCDHF}}^v$} 
		& \colhead{$\tau_{\rm{AS}}^l$} }
	\startdata
1  & $3s^{2}\,3p^{5}~^{2}P_{3/2}^{\circ}$                                           & 0        &  0     &     0       &    0     &      0       &   0        &   0          & 		     &          &         \\
2  & $3s^{2}\,3p^{5}~^{2}P_{1/2}^{\circ}$                                           & 15725    &  51    &     -42     &    -41.9 &      -42     &   -42      &   -76        & 1.43E-02&	1.43E-02&	1.44E-02\\
3  & $3s~^{2}S\,3p^{6}~^{2}S_{1/2}$                                                 & 289634   &  522   &     -398    &    -385  &      -398    &   -702     &   -4054      & 2.30E-10&	2.22E-10&	1.82E-10\\
4  & $3s^{2}\,3p^{4}(^{3}P)~^{3}P\,3d~^{4}D_{7/2}$                                  & 388485   &  -148  &     -919    &    224   &      223     &   329      &   731        & 1.73E-02&	1.73E-02&	1.77E-02\\
5  & $3s^{2}\,3p^{4}(^{3}P)~^{3}P\,3d~^{4}D_{5/2}$                                  & 388558   &  -77   &     -1094   &    151   &      156     &   274      &   713        & 1.66E-07&	2.05E-07&	3.00E-07\\
6  & $3s^{2}\,3p^{4}(^{3}P)~^{3}P\,3d~^{4}D_{3/2}$                                  & 389791   &  -84   &     -1266   &    259   &      228     &   289      &   731        & 1.67E-07&	1.63E-07&	2.26E-07\\
7  & $3s^{2}\,3p^{4}(^{3}P)~^{3}P\,3d~^{4}D_{1/2}$                                  & 391341   &  -65   &     213     &    214   &      213     &   276      &   685        & 1.18E-07&	1.04E-07&	1.40E-07\\
8  & $3s^{2}\,3p^{4}(^{3}P)~^{3}P\,3d~^{4}F_{9/2}$                                  & 417470   &  155   &     183     &    183   &      182     &   293      &   3998       & 9.00E-02&	8.99E-02&	8.91E-02\\
\enddata
\end{deluxetable*}

\clearpage

\startlongtable
\begin{deluxetable*}{clrrrrrrrr}
	\centering
	\tabletypesize{\small}
	\setlength{\tabcolsep}{3pt}
	\tablecaption{Excitation energies (in cm$^{-1}$) for the lower levels in \ion{Ni}{12}. $E_{\rm MCDHF}$ are our MCDHF
energies, $E_{\rm NIST}$ the NIST recommended values, and $E_{\rm Obs}$ our suggested experimental values (TN: tentative new identification; D identification as in 
\cite{delzanna_badnell:2016_ni_12}). $\Delta_1 =E_{\rm MCDHF}-E_{\rm NIST}$, $\Delta_2 =E_{\rm MCDHF}-E_{\rm Obs}$.
	\label{tab.ni_12}
}
	\tablehead{\colhead{Key} & \colhead{Level} & \colhead{$E_{\rm MCDHF}$} & \colhead{$E_{\rm NIST}$} & \colhead{$\Delta_1$} & \colhead{$E_{\rm Obs}$} &
       \colhead{$\Delta_2$}}
	\startdata
 1      &    $3s^{2}\,3p^{5}~^{2}P_{3/2}^{\circ}$                                          & 0        & 0       &  0       &  0          &   \\
 2      &    $3s^{2}\,3p^{5}~^{2}P_{1/2}^{\circ}$                                          & 23678    & 23629   &  49      &  23629    & 49 \\
 3      &    $3s~^{2}S\,3p^{6}~^{2}S_{1/2}$                                               & 339773   & 338615  &  1158    &  338615   & 1158 \\
 4      &    $3s^{2}\,3p^{4}(^{3}P)~^{3}P\,3d~^{4}D_{5/2}$                                 & 450451   &         &          &      &  \\   
 5      &    $3s^{2}\,3p^{4}(^{3}P)~^{3}P\,3d~^{4}D_{7/2}$                                 & 450941   & 454000  &  -3059   &  450812 TN  & -129 \\
 6      &    $3s^{2}\,3p^{4}(^{3}P)~^{3}P\,3d~^{4}D_{3/2}$                                 & 452175   &         &          &     & \\
 7      &    $3s^{2}\,3p^{4}(^{3}P)~^{3}P\,3d~^{4}D_{1/2}$                                 & 454809   &         &          &     & \\
 8      &    $3s^{2}\,3p^{4}(^{3}P)~^{3}P\,3d~^{4}F_{9/2}$                                 & 482542   & 485570  &  -3028   &  482378  TN & 164 \\
\enddata
\end{deluxetable*}

\clearpage

\startlongtable
\begin{deluxetable*}{crlrrrrrl}
	\centering
\tabletypesize{\tiny}
\setlength{\tabcolsep}{3pt}
	\tablecaption{
		Excitation energies (in cm$^{-1}$) and radiative lifetimes (in s) for the lowest 187 (196, 197, 193, 192, 194, 194) levels of  Cr~VIII (Mn~IX, Fe~X, Co~XI, Ni~XII, Cu~XIII, Zn~XIV). 
$E_{\rm MCDHF}$ - the present MCDHF calculations; $E_{\rm NIST}$  - observed energies listed in the NIST ASD \citep{Kramida.2018.V.p}; $\Delta E$: energy differences (in cm$^{-1}$)  between   the present calculations $E_{\rm MCDHF}$ and observed energies $E_{\rm NIST}$;
		$\tau_{\rm MCDHF}^l$ -  lifetimes in length form from the present  MCDHF  calculations;   $\tau_{\rm MCDHF}^v$ -  lifetimes  in velocity form from the present   MCDHF  calculations;  $LS$-composition -  the $LS$ eigenvector compositions. \label{tab.lev.all}}
	\tablehead{\colhead{$Z$} & \colhead{Key} & \colhead{Level}  & \colhead{$E_{\rm MCDHF}$}  & \colhead{$E_{\rm NIST}$} & \colhead{$\Delta E$} & \colhead{$\tau_{\rm MCDHF}^l$} & \colhead{$\tau_{\rm MCDHF}^v$} &$LS$-composition }
	\startdata
24   &    1      &    $3s^{2}\,3p^{5}~^{2}P_{3/2}^{\circ}$                                          & 0        & 0       &  0       &          &  	       &   0.95                                                                                                                                                                  \\       
24   &    2      &    $3s^{2}\,3p^{5}~^{2}P_{1/2}^{\circ}$                                          & 9927     & 9892    &  35      &  5.69E-02&	5.69E-02 &   0.95                                                                                                                                                                  \\       
24   &    3      &    $3s~^{2}S\,3p^{6}~^{2}S_{1/2}$                                                & 243659   & 242065  &  1594    &  3.26E-10&	3.07E-10 &   0.71 + 0.25~$3s^{2}\,3p^{4}(^{1}D)~^{1}D\,3d~^{2}S$                                                                                                                   \\       
24   &    4      &    $3s^{2}\,3p^{4}(^{3}P)~^{3}P\,3d~^{4}D_{7/2}$                                 & 328241   &         &          &  2.98E-02&	2.98E-02 &   0.95                                                                                                                                                                  \\       
24   &    5      &    $3s^{2}\,3p^{4}(^{3}P)~^{3}P\,3d~^{4}D_{5/2}$                                 & 328555   &         &          &  3.28E-07&	4.12E-07 &   0.94                                                                                                                                                                  \\       
24   &    6      &    $3s^{2}\,3p^{4}(^{3}P)~^{3}P\,3d~^{4}D_{3/2}$                                 & 329365   &         &          &  3.76E-07&	3.84E-07 &   0.94                                                                                                                                                                  \\
\enddata
	\tablecomments{ Table~\ref{tab.lev.all} is published in its entirety in the machine-readable format. The results for the levels, for which the NIST values are available, are shown here for guidance regarding its form and content.}
\end{deluxetable*}

\begin{deluxetable*}{lllllllllllll}
	\tabletypesize{\tiny}
	\setlength{\tabcolsep}{4pt}
	\tablecaption{Transition wavelengths $\lambda$ (in {\AA}), transition rates $A$ (in s$^{-1}$), weighted oscillator strengths $gf$, and line strengths $S$ (in a.u.) 
		between the states of  Cr~VIII (Mn~IX, Fe~X, Co~XI, Ni~XII, Cu~XIII, Zn~XIV) listed in Table~\ref{tab.lev.all}. E1 and E2 transition data are provided in both the length ($l$) and velocity ($v$) form. Type specifies the nature of electric (E) or magnetic (M) multipole transition, BF is the branching fraction from the upper level. The last column (Acc.) represents the estimated accuracies of the $S$ values using the NIST ASD terminology. Only transitions with BF  $\geq 10^{-5}$ are presented. \label{tab.trans.all} }
	\tablehead{\colhead{$Z$} & \colhead{$i$} & \colhead{$j$} &    
		\colhead{$\lambda$} & \colhead{Type}  &\colhead{BF} & \colhead{$A^l$} & \colhead{$gf^l$} & \colhead{$S^l$} & \colhead{$A^v$} & \colhead{$gf^v$} & \colhead{$S^v$}&  \colhead{Acc.}      }
	\startdata
26 & 1 &   2 &   6.3594E+03&	M1&	9.998E-01&	6.984E+01&	8.469E-07&	1.332E+00&					 &            &         &    \\
26 & 1 &   2 &   6.3594E+03&	E2&	2.156E-04&	1.506E-02&	1.826E-10&	2.798E-01&	1.455E-02&	1.764E-10	&2.702E-01&	D+ \\
26 & 1 &   3 &   3.4526E+02&	E1&	6.951E-01&	3.023E+09&	1.080E-01&	1.228E-01&	3.118E+09&	1.114E-01	&1.267E-01&	B+ \\
26 & 1 &   4 &   2.5741E+02&	M2&	1.000E+00&	5.777E+01&	4.591E-09&	3.503E+01&					 &            &         &    \\
26 & 1 &   5 &   2.5736E+02&	E1&	1.000E+00&	6.011E+06&	3.581E-04&	3.034E-04&	4.880E+06&	2.908E-04	&2.463E-04&	C  \\
26 & 1 &   6 &   2.5655E+02&	E1&	9.980E-01&	5.986E+06&	2.363E-04&	1.995E-04&	6.120E+06&	2.416E-04	&2.040E-04&	B  \\
26 & 1 &   7 &   2.5553E+02&	E1&	4.240E-01&	3.598E+06&	7.045E-05&	5.926E-05&	4.229E+06&	8.279E-05	&6.965E-05&	C+ \\
26 & 1 &   9 &   2.3820E+02&	E1&	1.573E-01&	3.496E+07&	5.948E-04&	4.664E-04&	3.656E+07&	6.220E-04	&4.877E-04&	B  \\
\enddata
	\tablecomments{ Table~\ref{tab.trans.all} is published in its entirety in the machine-readable format. Part of the values for $Z=26$ are shown here for guidance regarding its form and content.}
\end{deluxetable*}

\listofchanges
\end{document}